\documentclass[aip,apl,onecolumn,showpacs,reprint]{revtex4-1}

\usepackage{wrapfig}
\usepackage[]{graphicx,xcolor}
\usepackage{tabularx}
\usepackage{booktabs}
\usepackage{textcomp}
\usepackage{amsmath}
\usepackage{amssymb}
\usepackage[]{graphics}
\usepackage{amssymb}
\usepackage{amsmath}
\usepackage{color}
\usepackage{ifpdf}

\newcommand{\executeiffilenewer}[3]{%
\ifnum\pdfstrcmp{\pdffilemoddate{#1}}%
{\pdffilemoddate{#2}} > 0 {\immediate\write18{#3}}\fi}

\ifpdf\usepackage{epstopdf}\fi
\newcommand{\si}{SiO$_2$ }
\newcommand{\va}{VO$_2$ }

\begin{document}

\title{Thermal self-oscillations in radiative heat exchange}

\author{S. A. Dyakov}
\affiliation{Department of Materials and Nano Physics, School of Information and Communication Technology, KTH Royal Institute of Technology, Electrum 229, 16440 Kista, Sweden}

\author{J.~Dai}
\affiliation{Department of Materials and Nano Physics, School of Information and Communication Technology, KTH Royal Institute of Technology, Electrum 229, 16440 Kista, Sweden}

\author{M.~Yan}
\affiliation{Department of Materials and Nano Physics, School of Information and Communication Technology, KTH Royal Institute of Technology, Electrum 229, 16440 Kista, Sweden}

\author{M.~Qiu}
\affiliation{State Key Laboratory of Modern Optical Instrumentation, Department of Optical Engineering, Zhejiang University, 310027, Hangzhou, China}
\affiliation{Department of Materials and Nano Physics, School of Information and Communication Technology, KTH Royal Institute of Technology, Electrum 229, 16440 Kista, Sweden}
\date{January 8, 2015}

\begin{abstract}
We report the effect of relaxation-type self-induced temperature oscillations in the system of two parallel plates of \si and \va which exchange heat by thermal radiation in vacuum. The non-linear feedback in the self-oscillating system is provided by metal-insulator transition in VO$_2$. Using the method of fluctuational electrodynamics we show that under the action of an external laser of a constant power, the temperature of \va plate oscillates around its phase transition value. The period and amplitude of oscillations depend on the geometry of the structure. We found that at 500\,nm vacuum gap separating bulk \si plate and 50\,nm thick \va plate, the period of self-oscillations is 2\,s and the amplitude is 4\,K which is determined by phase switching at threshold temperatures of phase transition.
\end{abstract}


\maketitle 

A self-oscillation is undamped oscillation in dynamical system with non-linear feedback under the action of time-constant non-periodical external power source. The distinction of self-oscillations from forced oscillation is that the latter is driven by a source of power that is modulated externally.  The examples of self-oscillations are such natural phenomena as vibration of the plant leaves under the influence of a uniform air flow, formation of turbulent water flows in the river shallows, the voices of humans, animals and birds, heartbeat \cite{jenkins2013self}. The amplitude and waveform of self-oscillations are determined by nonlinear characteristics of the system. Two types of self-oscillations are commonly distinguished: harmonic-type and relaxation-type. In case of harmonic-type self-oscillation, the oscillatory system is capable for natural damping vibration on a resonance frequency. The shape of the waveform is close to sinusoidal. The relaxation-type self-oscillations has no resonant frequency. The actual period of oscillation depends on the switching at the thresholds, which fix the amplitude. The waveform of this type of self-oscillation may be very different from sinusoid \cite{jenkins2013self}.

A large variety of devices are based on electrical self-oscillations. Among them are integrated a.c. signal generators, inverters, pressure and temperature sensors, etc. The generation of electrical self-oscillations in devices which are based on \va was demonstrated in \cite{crunteanu2010voltage, beaumont2014current, leroy2012generation, lee2008metal, kim2010electrical, sakai2008high}.
The physical origin of self-oscillation in \cite{crunteanu2010voltage, beaumont2014current, leroy2012generation, lee2008metal, kim2010electrical, sakai2008high} is the Mott-like phase transition in \va at $T_{ph}=340$\,K \cite{qazilbash2007mott}. When the temperature of \va is smaller than $T_{ph}$ then it behaves as a uniaxial crystal with the optical axis orthogonal to its interfaces. On the other hand, when the temperature of \va is higher than $T_{ph}$, a \va plate is in its metallic phase and remains in this state for higher temperatures \cite{berglund1969electronic, PhysRevLett.17.1286}. The formation of the metallic phase occurs within $\sim$\,200 fs \cite{rini2005photoinduced}. Due to the phase transition, the current-voltage characteristics of VO$_2$-based devices possesses a negative differential resistance region that allows generating electrical self-oscillations \cite{crunteanu2010voltage, beaumont2014current, leroy2012generation, lee2008metal, kim2010electrical, sakai2008high}. Recently, it was shown that \va can be applied to the creation of thermal analog of such electronic components as transistor \cite{ben2014near}, diode \cite{ben2013phase} and memory \cite{dyakov2014near, PhysRevLett.113.074301, elzouka2014near}. In the above thermal  devices, the electric current is replaced by a radiative heat flow where the thermal photons play a role of heat transfer carriers.
\begin{figure}[b!]
\centering
\includegraphics[width=0.45\columnwidth]{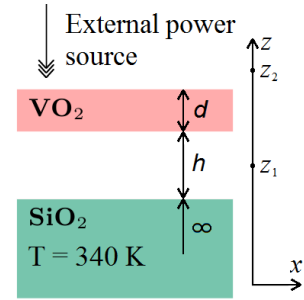}
\caption{The self-oscillating system consisting of thick \si plate and thin \va plate separated by a vacuum gap.The \va plate is illuminated by an external laser of a constant power.}
\label{fig1}
\end{figure}

In the present work we report the thermal analog of electrical self-oscillations in VO$_2$-based device. Thermal self-oscillations can be potentially interesting for practical realization of devices based on radiative heat flux control. A sketch of the self-oscillating system is shown in Fig.\,\ref{fig1}. The structure consists of two parallel plates of \si and \va which exchange thermal radiation through vacuum gap. We exclude such ways of heat transfer as convection and heat conduction via phonons and electrons. The choice of \si as a material of the first plate is based on a strong coupling between the \si and \va surface-phonon polaritons. The system is immersed in a thermal bath at temperature of $T_{bath}=300$\,K. We assume that the temperature of \si plate is fixed at 340\,K, while the temperature of \va is varied. The system is exposed by an external light source of constant power. We will show that at a certain power of the external light source, the metal-insulator phase transition in \va plays a role of positive non-linear feedback that ensures the self-induced oscillations of the temperature of \va plate. 

The temporal dynamics of \va plate temperature is described in terms of the standard energy balance equation \cite{PhysRevB.90.045414, PhysRevLett.113.074301}:
\begin{equation}
\label{intdiff}
\rho c_{v}d\frac{dT_2(t)}{dt} = F_{ext}(T_2) + F_{int}(T_2) + F_{bath}(T_2),
\end{equation}
where $\rho c_vd$ is thermal inertia of \va plate, with the mass density $\rho = 4.6$\,g/cm$^3$, the mass heat capacitance at constant volume $c_{v}$ and the thickness of \va plate $d$. In calculations, the temperature dependence of the mass heat capacitance of \va is accounted for by the Debye model with Debye temperature of 750\,K, the molar mass of 85.92\,g/mol and the number of atoms in \va molecule of 3. On the right-hand side of Eq.\,(\ref{intdiff}), the term $F_{ext}$ is a portion of the external power which is absorbed in VO$_2$. $F_{ext}$ is defined as $aF_0$, where $F_0$ is the laser power and $a$ is absorption coefficient of \va plate. The parameter $a$ can be calculated by the scattering matrix formalism \cite{ko88}. The term $F_{int}$ in Eq. (\ref{intdiff}) is the change of internal energy of \va plate during the time $dt$ due to the radiative heat exchange between plates:
\begin{equation}
\label{eq1} 
F_{int}(T_2) = F_{11}(T_2) - F_{12}(T_2) -F_{21}(T_2)-F_{22}(T_2),
\end{equation}
where $F_{ij}$ is the energy flux of thermal radiation of $i$-th plate in $z_j$ coordinate as shown in Fig.\,\ref{fig1}. Indices $i=1$ and 2 stand for \si and \va plates correspondingly; $z_1$ denotes any coordinate in the separation gap and $z_2$ denotes any coordinate in the upper semi-infinite vacuum. The radiative heat transfer between plates is calculated with a fluctuation dissipation theorem \cite{rytov1959theory}. The energy fluxes $F_{ij}$ are expressed as \cite{PhysRevA.84.042102}:
\begin{equation}
\label{eq1}
F_{ij} = \sum_{{s,p}}\int_0^{\infty}\frac{d\omega}{2\pi}\Theta(\omega, T_i)\int_0^\infty\frac{k_xdk_x}{2\pi}f_{ij}(\omega,k_x).
\end{equation}
where $\omega$ is the angular frequency and $k_x$ is the $x$-component of wavevector of thermal radiation from $i$-th plate; $f_{ij}(\omega, k_x)$ is the monochromatic flux of thermal radiation at certain $k_x$; $\Theta(\omega, T_i) = \hbar\omega/\left[exp(\hbar\omega/k_BT_i)\right]$ is the mean energy of Planck oscillator and $k_B$ is Boltzmann's constant. Expression (\ref{eq1}) accounts for contribution of both $s$- and $p$-polarizations. The monochromatic flux of thermal radiation $f_{ij}(\omega, k_x)$ is calculated using the complex amplitude reflectance and transmittance of the plates. For $k_x<\omega/c$, the coefficients $f_{ij}(\omega, k_x)$ are given by the following expressions:
\begin{align}
\label{eq21}f_{11}&= \left(1-|r_1|^2\right) \left(1-|r_2|^2\right)|D|^{-2},\\
f_{21}&= \left(1-|r_1|^2\right) \left(1-|r_2|^2-|t_2|^2\right)|D|^{-2},\\
f_{12}&= \left(1-|r_1|^2\right) |t_2|^2|D|^{-2},\\
f_{22}&= 1-|r_{02}|^2-\left(1-|r_{1}|^2\right)|t_{2}|^2|D|^{-2},
\end{align}
where $D=1-r_1r_2e^{2ik_{z0}h}$ is the Fabry-Perot like denominator and $h$ is the separation distance. For $k_x>\omega/c$
\begin{align}
f_{11}&= f_{21} = 4\mathrm{Im}(r_1)\mathrm{Im}(r_2)e^{-2|k_{z0}|h}|D|^{-2}\\
\label{eq22}f_{12}&= f_{22} = 0.
\end{align}
In expressions (\ref{eq21})--(\ref{eq22}), $r_i$ and $t_i$ are the complex amplitude reflectance and transmittance of the $i$-th plate, $r_{02}$ is the complex amplitude reflectance and  transmittance of the whole structure from the side of \va plate and $k_{z0}$ is the $z$-component of the wavevector in vacuum. Parameters $r_i$ and $t_i$ can be calculated by means of the scattering matrix method \cite{ko88, francoeur2009solution, PhysRevB.90.045414}. The Fresnel coefficients used in construction of the scattering matrices, accounting for anisotropy of \va plate, are described in \cite{ben2013phase, guo2014fluctuational}. The term $F_{bath}$ in Eq. (\ref{intdiff})  denotes the power which is absorbed in the \va plate due to the thermal bath.

\begin{figure}[b!]
\centering
\includegraphics[width=1\columnwidth]{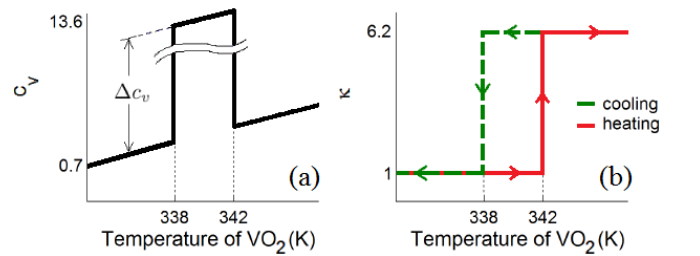}
\caption{Temperature dependence of the mass heat capacitance (a) and imaginary part of refractive index of \va at $\omega = 150$\,THz (b) near the phase transition temperature.}
\label{figcvkappa}
\end{figure}

The metal-insulator transition of \va is characterized by the hysteresis of optical constants and the latent heat which is absorbed in \va during the phase transformations. In calculations, we assume that the phase transition occurs in the temperature range $\{T_{ph}-\Delta T,T_{ph}+\Delta T\}$, where the heat capacitance equals to $c_v = c_{vo}+\Delta c_v$ as shown in Fig.\,\ref{figcvkappa}a. The term $c_{vo}$ is the heat capacitance of \va calculated by the Debye model while the term $\Delta c_v$ has a meaning an extra heat capacitance which is found from the expression for the latent heat: $L = 2\Delta c_v\Delta T$, where $\Delta T = 2$\,K and $L = 51.49$\,J/g \cite{berglund1969electronic}. The hysteresis of optical constants is modeled by an elementary non-ideal relay which is represented by rectangular loops on the $T_2-n$ and $T_2-\kappa$ coordinate planes. See example of such loop in Fig.\,\ref{figcvkappa}b for the extinction coefficient $\kappa$ at $\omega=150$\,THz. Hence, the metal-insulator transition is described by two temperatures, namely $T_{ph}\pm\Delta T$, which we call as threshold temperatures later on. As will be shown later, the presence of hysteresis is crucial for achieving self-oscillations.

\begin{figure}[t!]
\centering
\includegraphics[width=0.9\columnwidth]{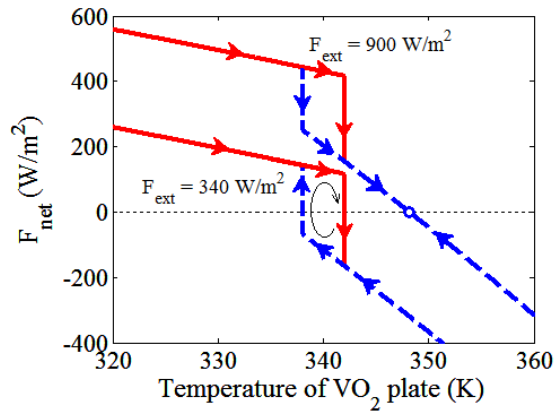}
\caption{The net power flux $F_{net}$ for \va plate as a function of its temperature, $T_2$. The arrows show the directions of temperature and phase transformations on the phase trajectories. Solid (dashed) lines denotes the regions of the phase diagram where \va is in its crystalline (metallic) state. Vacuum gap $h=50$\,nm, thickness of \va plate $d=500$\,nm, laser power $F_{0}=$\,340 and 900\,W/m$^2$.}
\label{fighyst}
\end{figure}

The phase portrait of the metal insulator transition in $d = 50$\,nm thick \va plate which is separated from \si by $h=500$\,nm vacuum gap is shown in Fig.\,\ref{fighyst} as a dependence of the net power flux for \va plate, $F_{net}\equiv F_{int}+F_{ext}+F_{bath}$, on the temperature $T_2$ for two different laser powers. The directions of temperature relaxations are determined by the sign of $F_{net}$ and are shown in Fig.\,\ref{fighyst} by arrows. Both $F_{net}(T_2)$ dependences possess the hysteresis which consists of the jumps of net power flux $F_{net}$ at the threshold temperatures $T_{ph}\pm \Delta T$ as well as of the relaxation regions between them. When $F_{ext}=900$\,W/m$^2$, the phase trajectory meets the x-axis  at 348.1\,K. It means that the system termalizes to 348.1\,K at the above external power flux. In the case $F_{ext}=340$\,W/m$^2$, the directions of temperature relaxation are opposite in metallic and insulator phases which causes the periodic alterations of \va plate temperature. Please note, that in the simulations described above, we assume that the derivatives $dn/dT_2$ and $d\kappa/dT_2$ if not infinite, are large enough at $T_2=T_{ph}\pm \Delta T$, which is a key factor causing the self-induced oscillations. Our additional simulations have shown, that due to thermal inertia of \va plate, the self-oscillations occur even if the parameters $n(T_2)$ and $\kappa(T_2)$ change not instantly but in a small temperature range $dT_2<1.6$\,K around threshold values $T_{ph}\pm \Delta T$. 

The jumps of net power flux $F_{net}$ is essentially non-linear process owing to the change of internal structure of \va during the phase transition. Note that the oscillations occur under a time-constant external power flux. Thus, the considered alterations are relaxation-type self-oscillations of \va plate temperature, where the metal-insulator phase transition plays a role of positive non-linear feedback. The oscillatory system never gets the stationary state but constantly thermalizes along its phase trajectories.
The similar jumps of the net heat flux at the threshold temperatures were studied in \cite{dyakov2014near} where $F_{net}(T_2)$ demonstrated the effect of negative differential thermal conductance from \si plate to \va plate. Unlike in \cite{dyakov2014near}, in the present work the thickness of plates and the separation distance between them were chosen in such a way that the dependence $F_{net}(T_2)$ is monotonically decreasing.
\begin{figure}[t!]
\centering
\includegraphics[width=0.89\columnwidth]{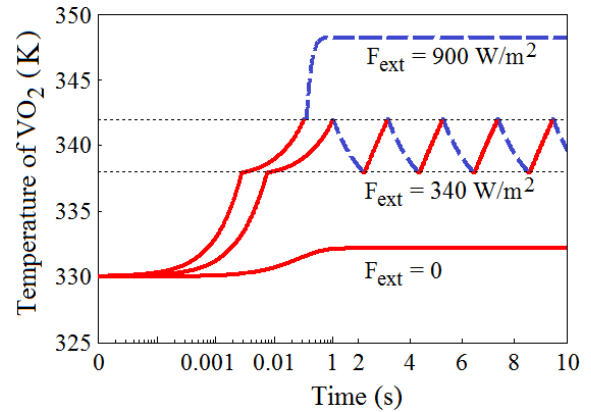}
\caption{The time evolutions of \va plate temperature at different external powers, $F_{ext}$. Solid (dashed) lines denotes the time intervals when \va is in its crystalline (metallic) phase.}
\label{fig11}
\end{figure}

In order to explicitly demonstrate the self-oscillation of \va temperature as well as to find its period, let us simulate the dynamics of the radiative heat exchange between the plates. We found the temporal dependencies of \va temperature by solving the integro-differential equation (\ref{intdiff}) with initial condition $T_2(0) = 330$\,K. Fig.\,\ref{fig11} shows the time evolutions $T_2(t)$ obtained for different external power fluxes. In the case of absence of external power flux ($F_{ext}=0$), the \va temperature increases to 333\,K without the phase transition. When $F_{ext}=900$\,W/m$^2$, the \va plate transits from insulator to metallic state and thermalizes to a constant temperature 348.1\,K that corresponds to the intersection point between the phase trajectory and the x-axis in Fig.\,\ref{fighyst}. In the time interval from 0 to 0.004\,s the temperature increases faster than in the subsequent interval from 0.004\,s to 0.38\,s. This is due to the fact that in temperature range 338--342\,K the energy which is absorbed in \va plate is spent for the phase transition. Finally, in the case of 340\,W/m$^2$ external power, the temperature $T_2$ starts to oscillate after 1\,s relaxation to 342\,K. The period of oscillation is about 2\,s and is determined by the thermal inertia of \va plate, separation distance between the plates, laser power, threshold temperatures and latent heat of the phase transition. The temperature oscillations are accompanied by periodical switches between insulator and metallic states of VO$_2$.

In conclusion, we have theoretically demonstrated the effect of relaxation-type temperature self-oscillation in radiative heat exchange between \va  and \si plates in vacuum. The hysteresis of optical constants of \va at threshold temperatures acts as non-linear feedback that support the self-oscillations. The oscillation period is determined by relaxation characteristics of the system and the threshold temperatures. Generalization of the studied phenomenon to the phonon- or electron-mediated heat exchange between different parts of the system may reveal the potential for significant shortening of the oscillation period. We believe that the discussed structure is a theoretical prototype of a thermal oscillations generator that could be useful for creation of devices based on radiative heat flux control.

This work is supported by the Swedish Research Council (VR) and VR's Linnaeus center in Advanced Optics and Photonics (ADOPT). M.Q. acknowledges the support by the National Natural Science Foundation of China (Grants Nos. 61275030, 61205030, and 61235007). S.D. acknowledges the Olle Erikssons Foundation for Materials Engineering for support.



%


\begin{thebibliography}{22}%
\makeatletter
\providecommand \@ifxundefined [1]{%
 \@ifx{#1\undefined}
}%
\providecommand \@ifnum [1]{%
 \ifnum #1\expandafter \@firstoftwo
 \else \expandafter \@secondoftwo
 \fi
}%
\providecommand \@ifx [1]{%
 \ifx #1\expandafter \@firstoftwo
 \else \expandafter \@secondoftwo
 \fi
}%
\providecommand \natexlab [1]{#1}%
\providecommand \enquote  [1]{``#1''}%
\providecommand \bibnamefont  [1]{#1}%
\providecommand \bibfnamefont [1]{#1}%
\providecommand \citenamefont [1]{#1}%
\providecommand \href@noop [0]{\@secondoftwo}%
\providecommand \href [0]{\begingroup \@sanitize@url \@href}%
\providecommand \@href[1]{\@@startlink{#1}\@@href}%
\providecommand \@@href[1]{\endgroup#1\@@endlink}%
\providecommand \@sanitize@url [0]{\catcode `\\12\catcode `\$12\catcode
  `\&12\catcode `\#12\catcode `\^12\catcode `\_12\catcode `\%12\relax}%
\providecommand \@@startlink[1]{}%
\providecommand \@@endlink[0]{}%
\providecommand \url  [0]{\begingroup\@sanitize@url \@url }%
\providecommand \@url [1]{\endgroup\@href {#1}{\urlprefix }}%
\providecommand \urlprefix  [0]{URL }%
\providecommand \Eprint [0]{\href }%
\providecommand \doibase [0]{http://dx.doi.org/}%
\providecommand \selectlanguage [0]{\@gobble}%
\providecommand \bibinfo  [0]{\@secondoftwo}%
\providecommand \bibfield  [0]{\@secondoftwo}%
\providecommand \translation [1]{[#1]}%
\providecommand \BibitemOpen [0]{}%
\providecommand \bibitemStop [0]{}%
\providecommand \bibitemNoStop [0]{.\EOS\space}%
\providecommand \EOS [0]{\spacefactor3000\relax}%
\providecommand \BibitemShut  [1]{\csname bibitem#1\endcsname}%
\let\auto@bib@innerbib\@empty
\bibitem [{\citenamefont {Jenkins}(2013)}]{jenkins2013self}%
  \BibitemOpen
  \bibfield  {author} {\bibinfo {author} {\bibfnamefont {A.}~\bibnamefont
  {Jenkins}},\ }\href@noop {} {\bibfield  {journal} {\bibinfo  {journal}
  {Physics Reports}\ }\textbf {\bibinfo {volume} {525}},\ \bibinfo {pages}
  {167} (\bibinfo {year} {2013})}\BibitemShut {NoStop}%
\bibitem [{\citenamefont {Crunteanu}\ \emph {et~al.}(2010)\citenamefont
  {Crunteanu}, \citenamefont {Givernaud}, \citenamefont {Leroy}, \citenamefont
  {Mardivirin}, \citenamefont {Champeaux}, \citenamefont {Orlianges},
  \citenamefont {Catherinot},\ and\ \citenamefont
  {Blondy}}]{crunteanu2010voltage}%
  \BibitemOpen
  \bibfield  {author} {\bibinfo {author} {\bibfnamefont {A.}~\bibnamefont
  {Crunteanu}}, \bibinfo {author} {\bibfnamefont {J.}~\bibnamefont
  {Givernaud}}, \bibinfo {author} {\bibfnamefont {J.}~\bibnamefont {Leroy}},
  \bibinfo {author} {\bibfnamefont {D.}~\bibnamefont {Mardivirin}}, \bibinfo
  {author} {\bibfnamefont {C.}~\bibnamefont {Champeaux}}, \bibinfo {author}
  {\bibfnamefont {J.-C.}\ \bibnamefont {Orlianges}}, \bibinfo {author}
  {\bibfnamefont {A.}~\bibnamefont {Catherinot}}, \ and\ \bibinfo {author}
  {\bibfnamefont {P.}~\bibnamefont {Blondy}},\ }\href@noop {} {\bibfield
  {journal} {\bibinfo  {journal} {Sci. Technol. Adv. Mater.}\ }\textbf
  {\bibinfo {volume} {11}},\ \bibinfo {pages} {065002} (\bibinfo {year}
  {2010})}\BibitemShut {NoStop}%
\bibitem [{\citenamefont {Beaumont}\ \emph {et~al.}(2014)\citenamefont
  {Beaumont}, \citenamefont {Leroy}, \citenamefont {Orlianges},\ and\
  \citenamefont {Crunteanu}}]{beaumont2014current}%
  \BibitemOpen
  \bibfield  {author} {\bibinfo {author} {\bibfnamefont {A.}~\bibnamefont
  {Beaumont}}, \bibinfo {author} {\bibfnamefont {J.}~\bibnamefont {Leroy}},
  \bibinfo {author} {\bibfnamefont {J.-C.}\ \bibnamefont {Orlianges}}, \ and\
  \bibinfo {author} {\bibfnamefont {A.}~\bibnamefont {Crunteanu}},\ }\href@noop
  {} {\bibfield  {journal} {\bibinfo  {journal} {J. Appl. Phys.}\ }\textbf
  {\bibinfo {volume} {115}},\ \bibinfo {pages} {154502} (\bibinfo {year}
  {2014})}\BibitemShut {NoStop}%
\bibitem [{\citenamefont {Leroy}\ \emph {et~al.}(2012)\citenamefont {Leroy},
  \citenamefont {Crunteanu}, \citenamefont {Givernaud}, \citenamefont
  {Orlianges}, \citenamefont {Champeaux}, \citenamefont {Blondy} \emph
  {et~al.}}]{leroy2012generation}%
  \BibitemOpen
  \bibfield  {author} {\bibinfo {author} {\bibfnamefont {J.}~\bibnamefont
  {Leroy}}, \bibinfo {author} {\bibfnamefont {A.}~\bibnamefont {Crunteanu}},
  \bibinfo {author} {\bibfnamefont {J.}~\bibnamefont {Givernaud}}, \bibinfo
  {author} {\bibfnamefont {J.-C.}\ \bibnamefont {Orlianges}}, \bibinfo {author}
  {\bibfnamefont {C.}~\bibnamefont {Champeaux}}, \bibinfo {author}
  {\bibfnamefont {P.}~\bibnamefont {Blondy}},  \emph {et~al.},\ }\href@noop {}
  {\bibfield  {journal} {\bibinfo  {journal} {International journal of
  microwave and wireless technologies}\ }\textbf {\bibinfo {volume} {4}}
  (\bibinfo {year} {2012})}\BibitemShut {NoStop}%
\bibitem [{\citenamefont {Lee}\ \emph {et~al.}(2008)\citenamefont {Lee},
  \citenamefont {Kim}, \citenamefont {Lim}, \citenamefont {Yun}, \citenamefont
  {Choi}, \citenamefont {Chae}, \citenamefont {Kim},\ and\ \citenamefont
  {Kim}}]{lee2008metal}%
  \BibitemOpen
  \bibfield  {author} {\bibinfo {author} {\bibfnamefont {Y.~W.}\ \bibnamefont
  {Lee}}, \bibinfo {author} {\bibfnamefont {B.-J.}\ \bibnamefont {Kim}},
  \bibinfo {author} {\bibfnamefont {J.-W.}\ \bibnamefont {Lim}}, \bibinfo
  {author} {\bibfnamefont {S.~J.}\ \bibnamefont {Yun}}, \bibinfo {author}
  {\bibfnamefont {S.}~\bibnamefont {Choi}}, \bibinfo {author} {\bibfnamefont
  {B.-G.}\ \bibnamefont {Chae}}, \bibinfo {author} {\bibfnamefont
  {G.}~\bibnamefont {Kim}}, \ and\ \bibinfo {author} {\bibfnamefont {H.-T.}\
  \bibnamefont {Kim}},\ }\href@noop {} {\bibfield  {journal} {\bibinfo
  {journal} {Appl. Phys. Lett.}\ }\textbf {\bibinfo {volume} {92}},\ \bibinfo
  {pages} {162903} (\bibinfo {year} {2008})}\BibitemShut {NoStop}%
\bibitem [{\citenamefont {Kim}\ \emph {et~al.}(2010)\citenamefont {Kim},
  \citenamefont {Kim}, \citenamefont {Choi}, \citenamefont {Chae},
  \citenamefont {Lee}, \citenamefont {Driscoll}, \citenamefont {Qazilbash},\
  and\ \citenamefont {Basov}}]{kim2010electrical}%
  \BibitemOpen
  \bibfield  {author} {\bibinfo {author} {\bibfnamefont {H.-T.}\ \bibnamefont
  {Kim}}, \bibinfo {author} {\bibfnamefont {B.-J.}\ \bibnamefont {Kim}},
  \bibinfo {author} {\bibfnamefont {S.}~\bibnamefont {Choi}}, \bibinfo {author}
  {\bibfnamefont {B.-G.}\ \bibnamefont {Chae}}, \bibinfo {author}
  {\bibfnamefont {Y.~W.}\ \bibnamefont {Lee}}, \bibinfo {author} {\bibfnamefont
  {T.}~\bibnamefont {Driscoll}}, \bibinfo {author} {\bibfnamefont {M.~M.}\
  \bibnamefont {Qazilbash}}, \ and\ \bibinfo {author} {\bibfnamefont
  {D.}~\bibnamefont {Basov}},\ }\href@noop {} {\bibfield  {journal} {\bibinfo
  {journal} {J. Appl. Phys.}\ }\textbf {\bibinfo {volume} {107}},\ \bibinfo
  {pages} {023702} (\bibinfo {year} {2010})}\BibitemShut {NoStop}%
\bibitem [{\citenamefont {Sakai}(2008)}]{sakai2008high}%
  \BibitemOpen
  \bibfield  {author} {\bibinfo {author} {\bibfnamefont {J.}~\bibnamefont
  {Sakai}},\ }\href@noop {} {\bibfield  {journal} {\bibinfo  {journal} {J.
  Appl. Phys.}\ }\textbf {\bibinfo {volume} {103}},\ \bibinfo {pages} {103708}
  (\bibinfo {year} {2008})}\BibitemShut {NoStop}%
\bibitem [{\citenamefont {Qazilbash}\ \emph {et~al.}(2007)\citenamefont
  {Qazilbash}, \citenamefont {Brehm}, \citenamefont {Chae}, \citenamefont {Ho},
  \citenamefont {Andreev}, \citenamefont {Kim}, \citenamefont {Yun},
  \citenamefont {Balatsky}, \citenamefont {Maple}, \citenamefont {Keilmann}
  \emph {et~al.}}]{qazilbash2007mott}%
  \BibitemOpen
  \bibfield  {author} {\bibinfo {author} {\bibfnamefont {M.~M.}\ \bibnamefont
  {Qazilbash}}, \bibinfo {author} {\bibfnamefont {M.}~\bibnamefont {Brehm}},
  \bibinfo {author} {\bibfnamefont {B.-G.}\ \bibnamefont {Chae}}, \bibinfo
  {author} {\bibfnamefont {P.-C.}\ \bibnamefont {Ho}}, \bibinfo {author}
  {\bibfnamefont {G.~O.}\ \bibnamefont {Andreev}}, \bibinfo {author}
  {\bibfnamefont {B.-J.}\ \bibnamefont {Kim}}, \bibinfo {author} {\bibfnamefont
  {S.~J.}\ \bibnamefont {Yun}}, \bibinfo {author} {\bibfnamefont
  {A.}~\bibnamefont {Balatsky}}, \bibinfo {author} {\bibfnamefont
  {M.}~\bibnamefont {Maple}}, \bibinfo {author} {\bibfnamefont
  {F.}~\bibnamefont {Keilmann}},  \emph {et~al.},\ }\href@noop {} {\bibfield
  {journal} {\bibinfo  {journal} {Science}\ }\textbf {\bibinfo {volume}
  {318}},\ \bibinfo {pages} {1750} (\bibinfo {year} {2007})}\BibitemShut
  {NoStop}%
\bibitem [{\citenamefont {Berglund}\ and\ \citenamefont
  {Guggenheim}(1969)}]{berglund1969electronic}%
  \BibitemOpen
  \bibfield  {author} {\bibinfo {author} {\bibfnamefont {C.}~\bibnamefont
  {Berglund}}\ and\ \bibinfo {author} {\bibfnamefont {H.}~\bibnamefont
  {Guggenheim}},\ }\href@noop {} {\bibfield  {journal} {\bibinfo  {journal}
  {Phys. Rev.}\ }\textbf {\bibinfo {volume} {185}},\ \bibinfo {pages} {1022}
  (\bibinfo {year} {1969})}\BibitemShut {NoStop}%
\bibitem [{\citenamefont {Barker}\ \emph {et~al.}(1966)\citenamefont {Barker},
  \citenamefont {Verleur},\ and\ \citenamefont
  {Guggenheim}}]{PhysRevLett.17.1286}%
  \BibitemOpen
  \bibfield  {author} {\bibinfo {author} {\bibfnamefont {A.~S.}\ \bibnamefont
  {Barker}}, \bibinfo {author} {\bibfnamefont {H.~W.}\ \bibnamefont {Verleur}},
  \ and\ \bibinfo {author} {\bibfnamefont {H.~J.}\ \bibnamefont {Guggenheim}},\
  }\href {\doibase 10.1103/PhysRevLett.17.1286} {\bibfield  {journal} {\bibinfo
   {journal} {Phys. Rev. Lett.}\ }\textbf {\bibinfo {volume} {17}},\ \bibinfo
  {pages} {1286} (\bibinfo {year} {1966})}\BibitemShut {NoStop}%
\bibitem [{\citenamefont {Rini}\ \emph {et~al.}(2005)\citenamefont {Rini},
  \citenamefont {Cavalleri}, \citenamefont {Schoenlein}, \citenamefont
  {L{\'o}pez}, \citenamefont {Feldman}, \citenamefont {Haglund~Jr},
  \citenamefont {Boatner}, \citenamefont {Haynes} \emph
  {et~al.}}]{rini2005photoinduced}%
  \BibitemOpen
  \bibfield  {author} {\bibinfo {author} {\bibfnamefont {M.}~\bibnamefont
  {Rini}}, \bibinfo {author} {\bibfnamefont {A.}~\bibnamefont {Cavalleri}},
  \bibinfo {author} {\bibfnamefont {R.~W.}\ \bibnamefont {Schoenlein}},
  \bibinfo {author} {\bibfnamefont {R.}~\bibnamefont {L{\'o}pez}}, \bibinfo
  {author} {\bibfnamefont {L.~C.}\ \bibnamefont {Feldman}}, \bibinfo {author}
  {\bibfnamefont {R.~F.}\ \bibnamefont {Haglund~Jr}}, \bibinfo {author}
  {\bibfnamefont {L.~A.}\ \bibnamefont {Boatner}}, \bibinfo {author}
  {\bibfnamefont {T.~E.}\ \bibnamefont {Haynes}},  \emph {et~al.},\ }\href@noop
  {} {\bibfield  {journal} {\bibinfo  {journal} {Optics letters}\ }\textbf
  {\bibinfo {volume} {30}},\ \bibinfo {pages} {558} (\bibinfo {year}
  {2005})}\BibitemShut {NoStop}%
\bibitem [{\citenamefont {Ben-Abdallah}\ and\ \citenamefont
  {Biehs}(2014)}]{ben2014near}%
  \BibitemOpen
  \bibfield  {author} {\bibinfo {author} {\bibfnamefont {P.}~\bibnamefont
  {Ben-Abdallah}}\ and\ \bibinfo {author} {\bibfnamefont {S.-A.}\ \bibnamefont
  {Biehs}},\ }\href@noop {} {\bibfield  {journal} {\bibinfo  {journal} {Phys.
  Rev. Lett.}\ }\textbf {\bibinfo {volume} {112}},\ \bibinfo {pages} {044301}
  (\bibinfo {year} {2014})}\BibitemShut {NoStop}%
\bibitem [{\citenamefont {Ben-Abdallah}\ and\ \citenamefont
  {Biehs}(2013)}]{ben2013phase}%
  \BibitemOpen
  \bibfield  {author} {\bibinfo {author} {\bibfnamefont {P.}~\bibnamefont
  {Ben-Abdallah}}\ and\ \bibinfo {author} {\bibfnamefont {S.-A.}\ \bibnamefont
  {Biehs}},\ }\href@noop {} {\bibfield  {journal} {\bibinfo  {journal} {Appl.
  Phys. Lett.}\ }\textbf {\bibinfo {volume} {103}},\ \bibinfo {pages} {191907}
  (\bibinfo {year} {2013})}\BibitemShut {NoStop}%
\bibitem [{\citenamefont {Dyakov}\ \emph
  {et~al.}(2014{\natexlab{a}})\citenamefont {Dyakov}, \citenamefont {Dai},
  \citenamefont {Yan},\ and\ \citenamefont {Qiu}}]{dyakov2014near}%
  \BibitemOpen
  \bibfield  {author} {\bibinfo {author} {\bibfnamefont {S.~A.}\ \bibnamefont
  {Dyakov}}, \bibinfo {author} {\bibfnamefont {J.}~\bibnamefont {Dai}},
  \bibinfo {author} {\bibfnamefont {M.}~\bibnamefont {Yan}}, \ and\ \bibinfo
  {author} {\bibfnamefont {M.}~\bibnamefont {Qiu}},\ }\href
  {http://arxiv.org/abs/1408.5831} {\bibfield  {journal} {\bibinfo  {journal}
  {arXiv preprint arXiv:1408.5831}\ } (\bibinfo {year}
  {2014}{\natexlab{a}})}\BibitemShut {NoStop}%
\bibitem [{\citenamefont {Kubytskyi}\ \emph {et~al.}(2014)\citenamefont
  {Kubytskyi}, \citenamefont {Biehs},\ and\ \citenamefont
  {Ben-Abdallah}}]{PhysRevLett.113.074301}%
  \BibitemOpen
  \bibfield  {author} {\bibinfo {author} {\bibfnamefont {V.}~\bibnamefont
  {Kubytskyi}}, \bibinfo {author} {\bibfnamefont {S.-A.}\ \bibnamefont
  {Biehs}}, \ and\ \bibinfo {author} {\bibfnamefont {P.}~\bibnamefont
  {Ben-Abdallah}},\ }\href {\doibase 10.1103/PhysRevLett.113.074301} {\bibfield
   {journal} {\bibinfo  {journal} {Phys. Rev. Lett.}\ }\textbf {\bibinfo
  {volume} {113}},\ \bibinfo {pages} {074301} (\bibinfo {year}
  {2014})}\BibitemShut {NoStop}%
\bibitem [{\citenamefont {Elzouka}\ and\ \citenamefont
  {Ndao}(2014)}]{elzouka2014near}%
  \BibitemOpen
  \bibfield  {author} {\bibinfo {author} {\bibfnamefont {M.}~\bibnamefont
  {Elzouka}}\ and\ \bibinfo {author} {\bibfnamefont {S.}~\bibnamefont {Ndao}},\
  }\href@noop {} {\bibfield  {journal} {\bibinfo  {journal} {Applied Physics
  Letters}\ }\textbf {\bibinfo {volume} {105}},\ \bibinfo {pages} {243510}
  (\bibinfo {year} {2014})}\BibitemShut {NoStop}%
\bibitem [{\citenamefont {Dyakov}\ \emph
  {et~al.}(2014{\natexlab{b}})\citenamefont {Dyakov}, \citenamefont {Dai},
  \citenamefont {Yan},\ and\ \citenamefont {Qiu}}]{PhysRevB.90.045414}%
  \BibitemOpen
  \bibfield  {author} {\bibinfo {author} {\bibfnamefont {S.~A.}\ \bibnamefont
  {Dyakov}}, \bibinfo {author} {\bibfnamefont {J.}~\bibnamefont {Dai}},
  \bibinfo {author} {\bibfnamefont {M.}~\bibnamefont {Yan}}, \ and\ \bibinfo
  {author} {\bibfnamefont {M.}~\bibnamefont {Qiu}},\ }\href {\doibase
  10.1103/PhysRevB.90.045414} {\bibfield  {journal} {\bibinfo  {journal} {Phys.
  Rev. B}\ }\textbf {\bibinfo {volume} {90}},\ \bibinfo {pages} {045414}
  (\bibinfo {year} {2014}{\natexlab{b}})}\BibitemShut {NoStop}%
\bibitem [{\citenamefont {Ko}\ and\ \citenamefont {Inkson}(1988)}]{ko88}%
  \BibitemOpen
  \bibfield  {author} {\bibinfo {author} {\bibfnamefont {D.~Y.~K.}\
  \bibnamefont {Ko}}\ and\ \bibinfo {author} {\bibfnamefont {J.}~\bibnamefont
  {Inkson}},\ }\href@noop {} {\bibfield  {journal} {\bibinfo  {journal}
  {Physical Review B}\ }\textbf {\bibinfo {volume} {38}},\ \bibinfo {pages}
  {9945} (\bibinfo {year} {1988})}\BibitemShut {NoStop}%
\bibitem [{\citenamefont {Rytov}(1959)}]{rytov1959theory}%
  \BibitemOpen
  \bibfield  {author} {\bibinfo {author} {\bibfnamefont {S.~M.}\ \bibnamefont
  {Rytov}},\ }\href@noop {} {\emph {\bibinfo {title} {Theory of electric
  fluctuations and thermal radiation}}},\ \bibinfo {type} {Tech. Rep.}\
  (\bibinfo  {institution} {DTIC Document},\ \bibinfo {year}
  {1959})\BibitemShut {NoStop}%
\bibitem [{\citenamefont {Messina}\ and\ \citenamefont
  {Antezza}(2011)}]{PhysRevA.84.042102}%
  \BibitemOpen
  \bibfield  {author} {\bibinfo {author} {\bibfnamefont {R.}~\bibnamefont
  {Messina}}\ and\ \bibinfo {author} {\bibfnamefont {M.}~\bibnamefont
  {Antezza}},\ }\href {\doibase 10.1103/PhysRevA.84.042102} {\bibfield
  {journal} {\bibinfo  {journal} {Phys. Rev. A}\ }\textbf {\bibinfo {volume}
  {84}},\ \bibinfo {pages} {042102} (\bibinfo {year} {2011})}\BibitemShut
  {NoStop}%
\bibitem [{\citenamefont {Francoeur}\ \emph {et~al.}(2009)\citenamefont
  {Francoeur}, \citenamefont {{Pinar Meng\"{u}\c{c}}},\ and\ \citenamefont
  {Vaillon}}]{francoeur2009solution}%
  \BibitemOpen
  \bibfield  {author} {\bibinfo {author} {\bibfnamefont {M.}~\bibnamefont
  {Francoeur}}, \bibinfo {author} {\bibfnamefont {M.}~\bibnamefont {{Pinar
  Meng\"{u}\c{c}}}}, \ and\ \bibinfo {author} {\bibfnamefont {R.}~\bibnamefont
  {Vaillon}},\ }\href@noop {} {\bibfield  {journal} {\bibinfo  {journal} {J.
  Quant. Spectrosc. Radiat. Transf.}\ }\textbf {\bibinfo {volume} {110}},\
  \bibinfo {pages} {2002} (\bibinfo {year} {2009})}\BibitemShut {NoStop}%
\bibitem [{\citenamefont {Guo}\ and\ \citenamefont
  {Jacob}(2014)}]{guo2014fluctuational}%
  \BibitemOpen
  \bibfield  {author} {\bibinfo {author} {\bibfnamefont {Y.}~\bibnamefont
  {Guo}}\ and\ \bibinfo {author} {\bibfnamefont {Z.}~\bibnamefont {Jacob}},\
  }\href@noop {} {\bibfield  {journal} {\bibinfo  {journal} {J. Appl. Phys.}\
  }\textbf {\bibinfo {volume} {115}},\ \bibinfo {pages} {234306} (\bibinfo
  {year} {2014})}\BibitemShut {NoStop}%
\end{thebibliography}
\end{document}